\documentclass{aa} 

\begin{document}


\title{Some like it cold: molecular emission and effective dust temperatures of dense cores in the Pipe Nebula}


\author{Jan Forbrich\inst{1,2} \and Karin \"Oberg\inst{2} \and
Charles J. Lada\inst{2} \and Marco Lombardi\inst{3} \and Alvaro Hacar\inst{1} \and Jo\~ao Alves\inst{1} \and Jill M. Rathborne\inst{4}}

\institute{Department of Astrophysics, University of Vienna, T\"urkenschanzstr. 17, 1180 Vienna, Austria \and Harvard-Smithsonian Center for Astrophysics, 60 Garden Street, Cambridge, MA 02138, USA \and University of Milan, Department of Physics, via Celoria 16, 20133, Milan, Italy \and CSIRO Astronomy and Space Science, Epping, Sydney, Australia}
 
\date{Received ; accepted}

\abstract
{}
{The Pipe Nebula is characterized by a low star formation rate, and is therefore an ideal environment to explore how initial conditions, including core characteristics, affect star formation efficiencies.
}
{In a continued study of the molecular core population of the Pipe Nebula, we present a molecular-line survey of 52 cores. Previous research has shown a variety of different chemical evolutionary stages among the cores. Using the Mopra radio telescope, we observed the ground rotational transitions of HCO$^{+}$, H$^{13}$CO$^{+}$, HCN, H$^{13}$CN, HNC, and N$_2$H$^{+}$. These data are complemented with near-infrared extinction maps to constrain the column densities, effective dust temperatures derived from \textit{Herschel} data, and NH$_3$-based gas kinetic temperatures. 
}
{The target cores are located across the nebula, span visual extinctions between 5 and 67 mag, and effective dust temperatures (averaged along the lines of sight) between 13 and 19~K. The extinction-normalized integrated line intensities, a proxy for the abundance in constant excitation conditions of optically thin lines, vary within an order of magnitude for a given molecule. The effective dust temperatures and gas kinetic temperatures are correlated, but the effective dust temperatures are consistently higher than the gas kinetic temperatures. Combining the molecular line and temperature data, we find that N$_2$H$^+$ is only detected toward the coldest and densest cores while other lines show no correlation with these core properties.
}
{Within this large sample, N$_2$H$^+$ is the only species to exclusively trace the coldest and densest cores, in agreement with chemical considerations. In contrast, the common high-density tracers HCN and HNC are present in a majority of cores, demonstrating the utility of these molecules to characterize cores over a large range of extinctions. The correlation between the effective dust temperatures and the gas kinetic temperatures suggests that the former are dominated by dust that is both dense and thermodynamically coupled to the dense gas traced by NH$_3$. A direct use of the effective dust temperatures in a determination of dust column densities from dust emission measurements would, however, result in an underestimate of the dust column densities.
}


\titlerunning{Some like it cold}
\authorrunning{Forbrich et al.}
\maketitle

\section{Introduction}

The Pipe Nebula with its population of dense low-mass cores is an ideal environment to study the initial conditions of star formation. The extinction maps of \citet{lom06} and \citet{rom10} have shown that the Pipe Nebula as a whole is similar to the $\rho$ Oph star-forming region in mass, size and distance ($d=130$~pc), and yet it has almost no active star formation \citep{for09,for10}. Using extinction mapping, C$^{18}$O, and NH$_3$ observations in particular, the core population has been characterized in detail. \citet{mue07} confirm the nature of the core population as consisting of distinct cores and not chance superpositions in the interstellar medium. In a study of the core mass function, as derived from extinction mapping, \citet{alv07} find that the core mass function is similar in shape to the stellar initial mass function from which it is different only by a scaling factor in mass. \citet{lad08} find that most of the cores are supported by thermal pressure. Additionally, \citet{rat08} present a survey of NH$_3$, CCS, and HC$_5$N emission toward some of the cores, and \citet{rat09} obtain an improved census of the core population. They conclude that there are 134 physically distinct dense cores associated with the Pipe Nebula. Also, they confirm the similarity in shape of the core mass function and the stellar initial mass function.

The large number and diverse set of well-characterized cores makes the Pipe Nebula a great laboratory to probe the connection between physical and chemical conditions in cores. Observationally, this can be done using a combination of dust extinction and emission maps, as well as molecular lines that trace dense molecular gas. Commonly used dense gas tracers can be divided into three categories: 1.\@ rare isotopologues of molecules with low dipole moments that are generally distributed and excited, 2.\@ molecules with high dipole moments that are generally distributed, but only excited at high densities, and 3.\@ molecules that are only observable in high-density cores because of chemical considerations. 

Rare isotopologues of CO are the most common probes that fall into the first category. These molecules will form at very low extinctions and have low critical densities, but because the lines can remain optically thin in high column density lines of sight, the emission will trace higher column density material, and may thus be good to simultaneously study cloud material and identify cores that span a large range of densities. HCN, HNC, HCO$^+$ and their isotopologues have higher critical densities and their emission should pinpoint higher density gas rather than diffuse emission from the surrounding lower density cloud. They should thus be useful to characterize core material at a range of core densities above a certain threshold ($\sim$10$^4$-10$^7$~cm$^{-3}$, depending on the excitation conditions of the molecule and transition). Finally N$_2$H$^+$, NH$_3$, and deuterated molecules are only expected in the coldest and densest cores because they rely on CO freeze-out to reach observable abundances. They should thus provide unique tracers of the coldest and densest cores \citep{ber01,ber02,taf06}. The goal of this study is to explore the molecular emission of a subset of these molecules towards a large sample of cores in a single star-forming region, the Pipe Nebula, to test the current ideas on the utility of these molecules as different kinds of core probes.

The molecular content in a sub-set of these cores has been explored before. Studying the evolutionary stages of various cores in the Pipe Nebula, \citet{fra10} presented molecular-line observations of four of these cores. Based on observations of 11 transitions of various molecules and a chemical-clock analysis, the authors conclude that the cores are all young but in different evolutionary stages. Subsequently, \citet{fra12a} extended this survey to a total of 14 sources and multiple transition lines. They find a rich chemistry and divide the sources into three groups -- diffuse, oxo-sulfurated and deuterated cores. The first group shows only few emission lines in the most abundant species. The second group, at higher column density, shows emission from oxo-sulfurated molecules while the third group, at the highest column densities, additionally shows emission from deuterated species. In a third paper, \citet{fra12b} extended the detailed molecular line survey of the first paper from four to a total of nine sources which again fall into three groups. The total sample of cores considered in these papers consists of 15 cores, as marked in Table~\ref{table_data}.

In this paper, we study chemical differences across the core population of the Pipe Nebula. We observed six molecular line transitions toward 52 out of 134 Pipe cores. In addition, we use near-infrared extinction maps and submillimeter data from the \textit{Herschel} satellite to derive effective dust temperatures. 
\section{Observations}

\subsection{Mopra millimeter--line data}

Observations were obtained in 2007 July 3--8 and 2008 July 20--25 at the 22-m Mopra radio telescope. They consist of position-switched single pointings of individual cloud cores in the Pipe Nebula with positions identified from extinction maps \citep{alv07,rat09} as listed in Table~\ref{table_data}. The 8 GHz~MOPS spectrometer was used to simultaneously observe the ground transitions of HCN, H$^{13}$CN, HCO$^{+}$, H$^{13}$CO$^{+}$, HNC, and N$_2$H$^{+}$, covering a frequency range of about 86--93 GHz. Each of these lines was covered with one spectral window each, 137.5 MHz wide with 4096 channels. This produced a velocity resolution of about 0.11~km\,s$^{-1}$. The FWHM Mopra beam size is $\sim$36$''$ \citep{lad05} at a frequency of 86~GHz, corresponding to an absolute scale of 0.02~pc at the distance of the Pipe Nebula (130 pc, \citealp{lom06}). The telescope pointing was checked approximately every hour using a suitably bright, nearby maser. 

The spectra were reduced using the ATNF Spectral Analysis Package (ASAP). All spectra are in $T_A^*$ with a median noise of 0.05~K per channel. We have searched all spectra for detections at about the system velocity and with more than one channel above the 3$\sigma$ threshold. Integrated intensities (in $T_A^*$) were obtained from the spectra, including hyperfine components where applicable. Corresponding error bars were conservatively calculated over the full line widths using error propagation, assuming an error for every spectral bin that is equivalent to the rms noise of the spectrum. The numerical values of errors and upper limits vary due to differences in integration time and weather conditions.

\subsection{Herschel submillimeter continuum data}

In addition to the Mopra observations we make use of an effective dust temperature map that was generated from \textit{Herschel} multi-band submillimeter observations, as calibrated with \textit{Planck} and extinction data. These effective dust temperatures are based on fits of the spectral energy distribution with a modified black body and represent weighted averages of the dust temperatures along the lines of sight, where warmer dust is weighted more than colder dust because the emission is stronger. The procedure is explained in detail in \citet{lom14}. Briefly, the optical depth, temperature, and spectral index maps from the {\it Planck} mission are used to predict and correct the {\it Herschel} fluxes in the various bands to obtain absolute calibration information. Optical depth and effective dust temperature maps are then obtained in gray-body fits with these two parameters while assuming a local value for the dust spectral index $\beta$ as taken from {\it Planck}/IRAS maps. Finally, the submillimeter opacity is tied to a near-infrared extinction map in a linear fit to obtain a column density map from the {\it Herschel} data. The angular resolution of the resulting column density and temperature maps is 36$''$ FWHM. While we can derive nominal temperature uncertainties between 0.06 K and 0.40 K, as listed in Table~\ref{table_data}, the total error budget is likely dominated by systematics. These include temperature or $\beta$ gradients along the line of sight, local spatial changes of $\beta$, other changes in the extinction law, or inaccuracies in the absolute flux calibration of {\it Herschel}. While these are hard to quantify, we point out that we are mainly using relative temperature measurements here, minimizing the effect of these systematics.

\begin{figure*}
\centering
\includegraphics[width=0.85\linewidth]{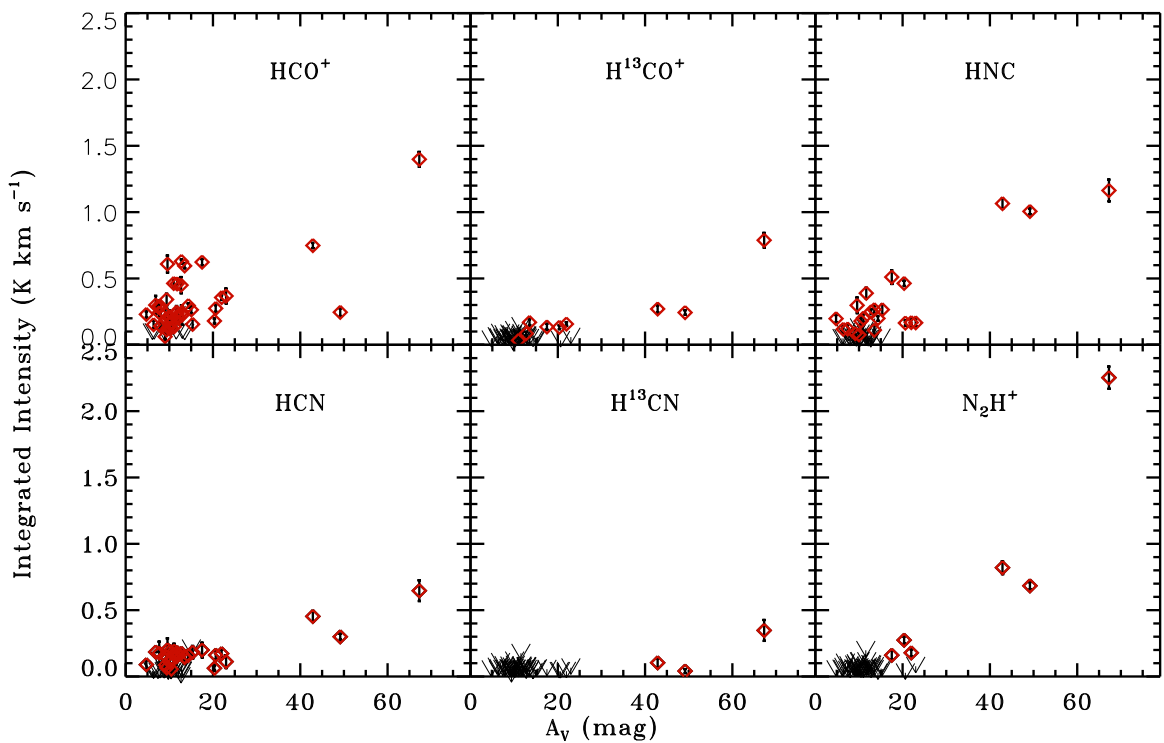}
\caption{Correlation of $\int T_A^*$ of each transition with $A_V$, including upper limits.\label{fig_pipe_corrN1}}
\end{figure*}

\begin{figure*}
\centering
\includegraphics[width=0.85\linewidth]{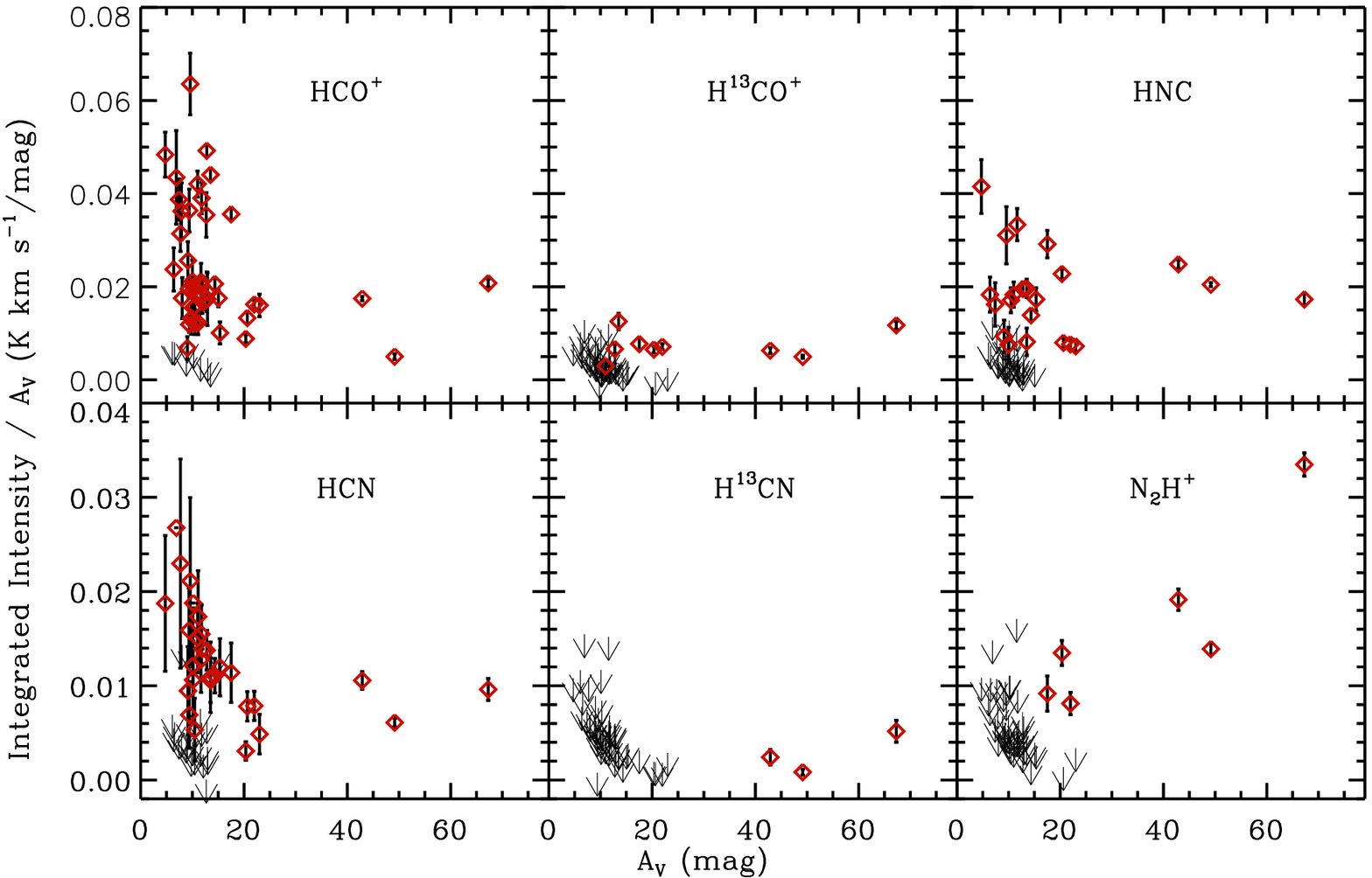}
\caption{Correlation of $\int T_A^*/A_V$ of each transition with $A_V$, including upper limits.\label{fig_pipe_corrN}}
\end{figure*}

\section{Results and Discussion}

\begin{table*}
\caption{Integrated line intensities and core temperatures}             
\label{table_data}      
\centering 
\small         
\begin{tabular}{rrrrllllllr}
\hline\hline       
Core\tablefootmark{a} & $\ell$    & $b$        & $A_V$ & N$_2$H$^+$      & HCN & H$^{13}$CN & HNC & HCO$^{+}$ & H$^{13}$CO$^{+}$ & $T_{\rm eff}{(\rm dust)}$ \\
                      & ($^\circ$)& ($^\circ$) & (mag) & \multicolumn{6}{c}{(K\,km\,s$^{-1}$)} &  (K)\\
\hline  		  
3\tablefootmark{b}  & -3.03  &  7.28 & 22.0   &0.18(3) &0.17(3) &       &0.17(2)  &0.36(2)  &0.16(1)  & 14.55(13)\\
4  & -2.99  &  7.01 & 12.8   &       &0.18(3) &        &0.25(2)  &0.63(2)  &0.09(1)  & 15.31(12)\\
5\tablefootmark{b}  & -2.97  &  6.85 & 11.9   &        &0.17(5) &        &	 &0.22(5)  &	     & 15.48(14)\\
7  & -2.95  &  7.25 &  9.6   &       &0.20(9) &        &0.30(6)  &0.61(6)  &	 & 15.80(14)\\
9\tablefootmark{b}  & -2.93  &  7.12 & 67.3   &2.25(8) &0.65(8) &0.35(8) &1.16(8)  &1.40(6)  &0.79(6)  & 13.50(21)\\
10 & -2.83  &  7.35 &  4.8\tablefootmark{c}& &0.09(3) & &0.20(3)  &0.23(2)  &	     & 17.51(21)\\
11\tablefootmark{b} & -2.71  &  6.96 & 11.0   &       &0.17(4) &        &0.20(3)  &0.46(3)  &0.03(1)  & 14.74(11)\\
12 & -2.68  &  6.78 &  6.9\tablefootmark{c}& &0.18(5) & &	  &0.30(7)  &	     & 16.55(19)\\
16\tablefootmark{b} & -2.54  &  6.35 & 11.7   &       &0.15(4) &        &0.39(4)  &0.25(5)  &	     & 15.87(12)\\
17 & -2.42  &  6.51 &  7.4   &       &        &        &0.12(4)  &0.29(3)  &	     & 17.44(13)\\
18 & -2.38  &  6.22 &  9.0   &       &        &        &	 &	   &	     & 16.90(13)\\
23 & -1.84  &  6.30 &  9.0   &        &   &	   &	     &0.06(2)  &	 & 16.85(13)\\
28 & -1.46  &  5.47 &  6.4   &       &        &        &0.12(2)  &0.15(3)  &	     & 17.77(26)\\
29\tablefootmark{b} & -1.43  &  5.90 & 13.5   &       &0.15(5) &        &0.11(4)  &	   &	     & 15.74(13)\\
30 & -1.41  &  5.75 & 10.5   &       &0.06(4) &        &0.18(3)  &0.17(3)  &	     & 16.53(13)\\
33 & -1.28  &  6.04 & 12.7   &       &        &        &	 &0.45(6)  &	     & 15.54(11)\\
34\tablefootmark{b} & -1.21  &  5.64 & 20.3   &0.27(3) &0.06(2) &       &0.46(2)  &0.18(2)  &0.13(1)  & 14.26(07)\\
35 & -1.19  &  5.26 & 20.6   &       &0.16(3) &        &0.16(2)  &0.27(3)  &	     & 15.51(08)\\
38 & -0.52  &  5.24 &  6.5   &       &        &        &	 &	   &	     & 17.45(17)\\
39\tablefootmark{b} & -0.50  &  4.44 & 10.0   &       &        &        &0.08(4)  &0.21(5)  &	     & 16.44(15)\\
40\tablefootmark{b} & -0.49  &  4.85 & 10.1   &       &0.11(4) &        &	 &0.20(2)  &	     & 16.59(13)\\
43 & -0.31  &  4.58 &  7.9   &        &   &	   &	     &0.29(5)  &	 & 17.05(16)\\
46\tablefootmark{b} &  0.07  &  4.62 & 12.9   &        &   &	   &	     &0.22(7)  &	 & 16.18(16)\\
47 &  0.08  &  3.86 &  8.8   &        &   &	   &	     &         &	 & 18.32(19)\\
51 &  0.23  &  4.55 & 10.4   &       &        &        &	 &0.13(3)  &	     & 16.71(13)\\
52 &  0.31  &  3.87 &  9.8   &       &        &        &	 &0.13(3)  &	     & 17.57(14)\\
53\tablefootmark{b} &  0.37  &  3.97 & 12.1   &       &        &        &	 &0.20(2)  &	     & 16.64(15)\\
55 &  0.53  &  4.78 &  9.1   &       &0.09(4) &        &0.09(3)  &0.23(4)  &	     & 16.73(15)\\
56 &  0.59  &  4.48 &  8.0   &       &        &        &	 &0.14(4)  &	     & 17.60(22)\\
60\tablefootmark{b} &  0.73  &  3.87 & 13.0   &        &   &	   &	     &         &	 & 16.48(11)\\
68 &  1.20  &  3.57 &  9.4   &        &0.07(3) &   &	     &0.11(2)  &	 & 18.33(17)\\
71\tablefootmark{b} &  1.31  &  3.76 & 49.2   &0.68(3) &0.30(2) &0.04(2) &1.01(2)  &0.25(3)  &0.24(2)  & 13.18(06)\\
72 &  1.33  &  3.93 & 11.6   &       &        &        &	 &	   &	     & 17.03(11)\\
74 &  1.38  &  4.40 & 13.5   &        &0.14(3) &   &0.27(3)  &0.60(2)  &0.17(2)  & 16.41(13)\\
76 &  1.41  &  3.71 & 14.4   &        &0.16(3) &   &0.20(2)  &0.30(2)  &	 & 16.23(13)\\
81 &  1.47  &  4.10 & 15.3   &        &0.18(5) &   &0.27(4)  &0.16(4)  &	 & 17.15(10)\\
82 &  1.48  &  3.79 & 15.1   &        &   &	   &	     &0.26(3)  &	 & 16.91(11)\\
84 &  1.51  &  6.41 &  7.7   &        &0.18(9) &   &	     &0.24(3)  &	 & 17.64(40)\\
86 &  1.52  &  7.08 & 17.5   &0.16(3) &0.20(6) &   &0.51(5)  &0.62(2)  &0.14(2)  & 14.32(19)\\
87\tablefootmark{b} &  1.52  &  3.92 & 23.0   &        &0.11(5) &   &0.17(3)  &0.37(6)  &	 & 15.86(08)\\
90 &  1.58  &  6.43 &  9.4   &        &   &	   &	     &0.34(4)  &	 & 16.95(20)\\
91 &  1.58  &  6.49 & 11.8   &        &   &	   &	     &0.46(4)  &	 & 16.78(19)\\
96\tablefootmark{b} &  1.71  &  3.65 & 42.9   &0.82(5) &0.45(4) &0.10(4) &1.07(4)  &0.75(2)  &0.27(3)  & 13.76(07)\\
99 &  1.77  &  6.93 & 10.1   &       &0.12(6) &        &	 &0.16(4)  &	     & 16.56(16)\\
100&  1.77  &  6.98 &  9.3   &       &0.15(4) &        &	 &0.18(2)  &	     & 16.67(16)\\
101&  1.80  &  3.88 & 10.3   &       &        &        &	 &	   &	     & 18.38(16)\\
102&  1.80  &  7.15 &  6.1   &       &        &        &	 &	   &	     & 18.29(28)\\
103&  1.85  &  3.76 & 11.8   &       &0.18(4) &        &	 &0.24(2)  &	     & 17.37(13)\\
110&  2.00  &  3.63 & 12.8   &        &   &	   &	     &0.24(5)  &	 & 17.99(14)\\
115&  2.13  &  3.55 & 11.6   &        &   &	   &	     &         &	 & 18.30(15)\\
117&  2.22  &  3.35 & 11.1   &        &0.19(5) &   &	     &0.14(3)  &	 & 18.39(12)\\
118&  2.22  &  3.41 & 10.2   &        &0.19(5) &   &	     &0.20(3)  &	 & 17.96(15)\\
\hline  									 \end{tabular}
\tablefoot{
1$\sigma$ errors in the last digit(s) shown in brackets (see text for a discussion of the temperature errors). Empty fields denote nondetections.\\
\tablefoottext{a}{Core numbers are from \citet{rat09}.}
\tablefoottext{b}{Cores also discussed by \citet{fra10,fra12a,fra12b}, but using core numbers from \citet{rat08} instead of \citet{rat09}. The corresponding numbers are as follows (Frau numbers in brackets): 3(6), 5(8), 9(12), 11(14), 16(20), 29(33), 34(40), 39(47), 40(48), 46(56), 53(65), 60(74), 71(87), 87(102), and 96(109).}
\tablefoottext{c}{Extinction data from \citet{lom06} instead of \citet{rom10}.}
}
\end{table*}

\subsection{Line intensities across the Pipe Nebula core population}

Integrated line intensities depend on a combination of total column density, molecular abundances, and excitation conditions (typically dominated by temperature and density). As a proxy for molecular abundances we have divided the line intensities by the visual extinction in each line of sight (similar to \citealp{fra12a}), providing extinction normalized line intensities that for optically thin lines should only depend on excitation and chemistry. For each core the integrated intensities are divided by the value of the corresponding pixel in the extinction maps of \citet{rom10}, or, in two cases\footnote{marked in Table~\ref{table_data}}, \citet{lom06} with its wider coverage. These extinction maps have an angular resolution of 20$''$ and 60$''$ FWHM, respectively.

The two extinction values separating the three groups of cores in the Frau et al. papers are $A_V\sim22.5$~mag and $A_V\sim15.0$~mag. Applying these numbers to our sample, we have four sources in the highest-extinction group, six sources in the second group, and 43 sources in the lowest-extinction group. Note that all 15 sources in the Frau et al. papers also appear in our sample. 
The two most frequently detected lines in our sample, above 3$\sigma$, are HCO$^+$ and HCN with 43 and 29 detections, respectively. In their isotopologues H$^{13}$CO$^+$ and H$^{13}$CN, there are only 9 and 3 detections, respectively. Finally, we detect HNC toward 22 and N$_2$H$^+$ toward six cores.

Our results are listed in Table~\ref{table_data}, and we show a plot of integrated line intensities over extinction in Fig.~\ref{fig_pipe_corrN1}. The three cores with the highest extinction levels ($A_V>40$~mag), namely Barnard 59, the Pipe Molecular Ring, and FeSt 1-457, are detected in all lines while many of the less extincted cores have upper limits in some or all of the transitions. HCN, HNC, and HCO$^{+}$ are detected across the full range of extinctions in our sample, with several non-detections at low extinction levels ($A_V<15$~mag). H$^{13}$CN and to a lesser degree H$^{13}$CO$^{+}$ are only detected toward the highest extinctions. N$_2$H$^{+}$ is detected only toward cores with high visual extinction of $A_V>17$~mag. Two of the high extinction cores do not display N$_2$H$^+$ emission, however, and we will discuss those below after we introduce the core temperatures.

The extinction-normalized integrated line intensities as a function of extinction are shown in Fig.~\ref{fig_pipe_corrN}. HCO$^+$ and HCN show a large scatter at low extinctions ($A_V<20$~mag) around means that are higher than for the high extinction cores, i.e., there is a marginal decrease. Some of the scatter is due to noise, but some may also be due to abundance variations and different excitation conditions in the lower density cores. 
The decreasing normalized line intensities with extinction could be due to a decreasing abundance but here these are instead indicative of lines that are becoming increasingly optically thick, thus no longer tracing the total column. Indeed, the intensity ratios of the isotopologues are far lower than what would be expected for optically thin transitions of HCO$^+$ and HCN. This result is supported by the flat trends of H$^{13}$CO$^+$ and H$^{13}$CN with extinction, which is expected for optically thin lines that are present at a constant abundance across the full range of core properties. While for HCO$^+$, the decrease of the extinction-normalized integrated intensities could partially also be due to depletion, it is thus more likely due to an optical depth effect since the decrease is not confirmed by the limited sample of H$^{13}$CO$^+$ detections. However, deeper observations of H$^{13}$CO$^+$ at the lowermost extinctions would be required for a definitive test.

N$_2$H$^+$ is the only molecule whose normalized emission clearly increases with extinction. For visual extinctions above $A_V=15$~mag, increasing abundances for N$_2$H$^+$ with extinction were already pointed out by \citet{ber01}. Since the N$_2$H$^+$ critical density is similar to those of HCN and HNC this may have a chemical rather than an excitation origin, particularly when considering that the excitation temperature of N$_2$H$^+$ in dense cores has been found to be fairly constant \citep{cas02}.

\subsection{Core properties as a function of temperature}

\begin{figure}
\centering
\includegraphics[bb= 115 15 415 310, scale=0.8]{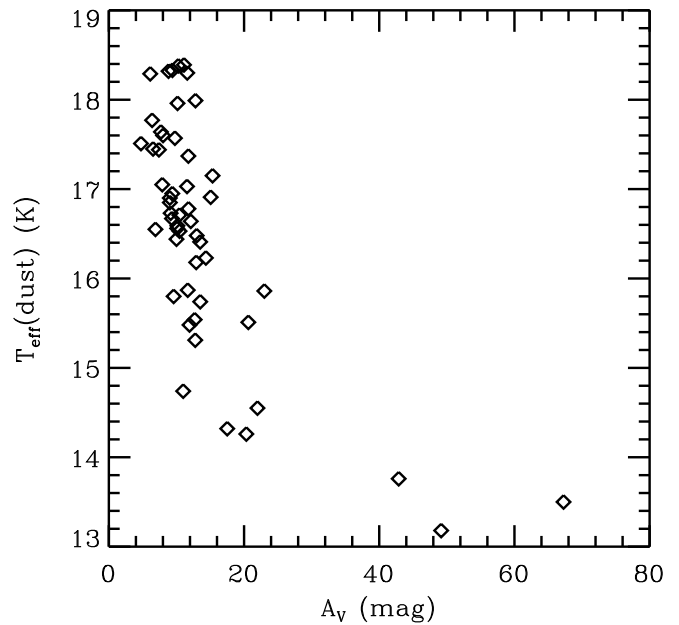}
\caption{Correlation of the effective dust temperature $T$ (K) and the core visual extinction $A_V$ (mag).\label{fig_TvsAV}}
\end{figure}

\begin{figure}
\centering
\includegraphics[bb= 115 15 415 310, scale=0.8]{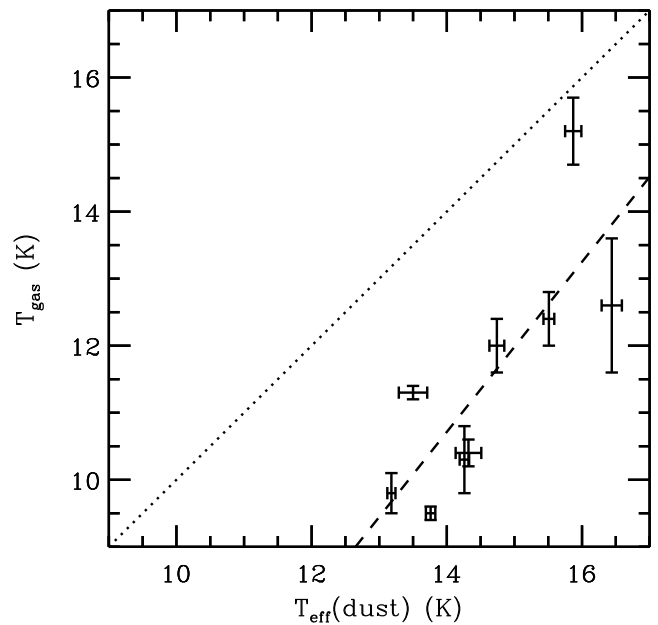}
\caption{Correlation of $T_K$ from NH$_3$ measurements reported by \citet{rat08} and the effective dust temperatures derived from \textit{Herschel} data. The gas temperatures from NH$_3$ are always lower than the effective dust temperatures. The dashed line shows a linear fit, and the dotted line indicates a direct correspondence of the two temperatures.\label{fig_T_hp_nh3}}
\end{figure}

The effective dust temperatures from \textit{Herschel} allow us to better understand the chemistry of these cores. Individual temperatures were extracted for each core at its observed position. It becomes apparent immediately that there is a strong anti-correlation between the core column density, as traced by the extinction map, and the effective dust temperature (Fig.~\ref{fig_TvsAV}). This indicates that the cores with the highest levels of extinction do contain the coldest material. The high column densities are thus not a consequence of a long column of warm diffuse emission; instead, the high column densities correspond to spatially concentrated high volume densities, shielding the inner regions of the cores from heating due to external radiation. Note that unlike previous studies of individual cores (e.g., \citealp{war02}, \citealp{gon04}), the data points shown in Fig.~\ref{fig_TvsAV} represent distinct cores and not a single coherent region. As a result, a direct comparison may not be particularly meaningful.

Since the effective dust temperatures may be biased due to the line-of-sight averaging discussed above, we compare them to gas kinetic temperatures derived from dense-gas--tracing NH$_3$ observations. Assuming thermodynamic coupling of dense dust and gas, the two quantities should be correlated, and if the dust emission is dominated by the core, the two temperatures should be almost identical. \citet{rat08} present measurements of the NH$_3$ gas temperatures for some of the Pipe cores, and in Fig.~\ref{fig_T_hp_nh3}, we show a direct comparison of the gas kinetic temperature $T_{\rm gas}$ from \citet{rat08} and the corresponding effective dust temperature. Figure~\ref{fig_T_hp_nh3} shows that while there is a direct linear correlation between the two quantities (dashed line), the effective dust temperatures are consistently higher than the gas temperatures, as could be expected from line-of-sight averaging of the dust that is not only tracing the densest and coldest parts of the clouds where dust and gas temperatures would be more closely coupled. Additionally, the dynamic range in the NH$_3$ temperatures is almost twice that of the effective dust temperatures, although that difference is mostly due to a single data point that also has the largest offset from the linear fit. 

Nevertheless, the linear relation between the two sets of temperature measurements underlines that the effective dust temperatures are physically meaningful for our core sample and dominated by dust that is thermodynamically coupled to the dense gas. Such coupling is expected for volume densities above 10$^{5..6}$~cm$^{-3}$ \citep{bur83,gol01}. While these densities are higher than those derived from the NH$_3$ observations, which are on the order of 10$^{4}$~cm$^{-3}$ \citep{rat08}, it seems plausible that higher densities could be reached in the interior of these cores. Thus, while the meaning of an absolute effective dust temperature is not obvious because of the line of sight averaging, the use of effective dust temperatures for relative comparisons within the core sample presented here appears to be justified. A direct use of the effective dust temperatures in a derivation of the dust column densities from dust emission measurements would, however, result in an underestimate of the dust column densities.

\begin{figure*}
\centering
\includegraphics[width=0.85\linewidth]{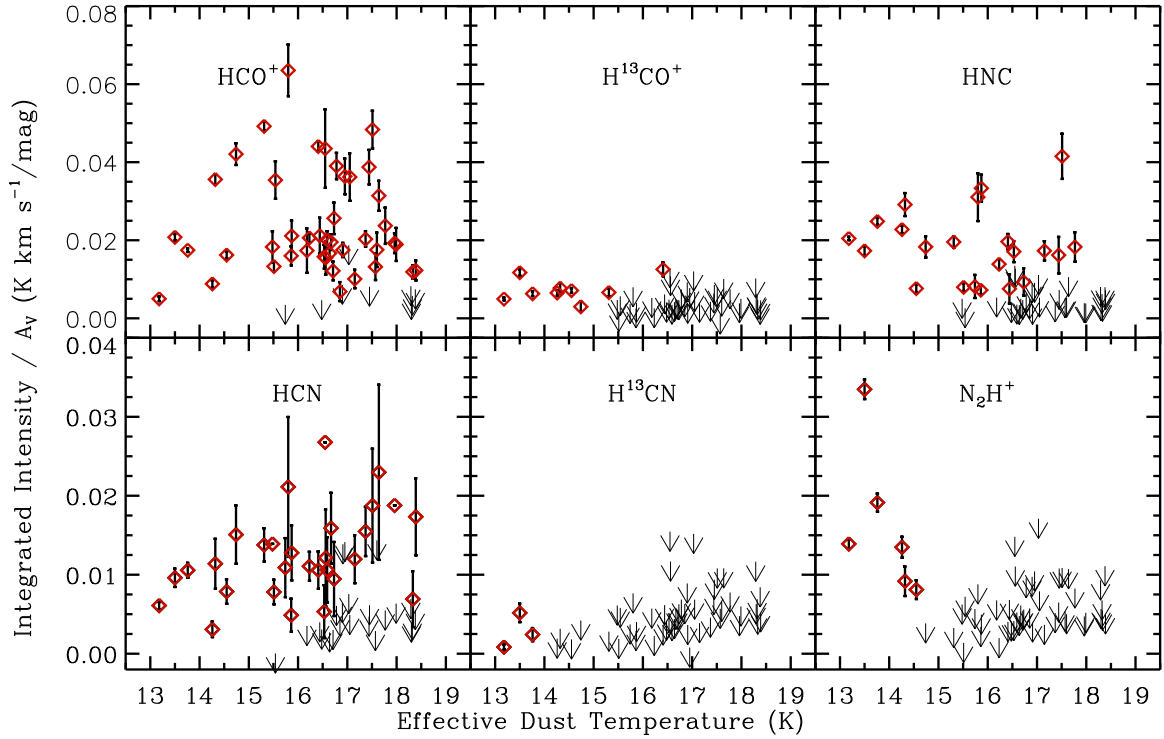}
\caption{Correlation of $\int T_A^*/A_V$ of each transition with temperature, including upper limits.\label{fig_pipe_corrN_wT2}}
\end{figure*}

In Fig.~\ref{fig_pipe_corrN1}, we show the integrated intensities in all six transitions as a function of column density, as traced by the extinction maps in $A_V$ (mag). All integrated line intensities increase with extinction, but the rate of increase differs by molecular transition. Additionally, in Fig.~\ref{fig_pipe_corrN_wT2}, we plot the extinction-normalized integrated line intensities as a function of the effective dust temperatures of the cores. The effective dust temperatures range from about 13~K to 19~K. Interestingly, the coldest cores are detected in all six transitions while that is not the case for the warmer cores. The trend is particularly clear in N$_2$H$^+$ where only the coldest cores of the sample have been detected. They have temperatures between 13.2~K and 14.6~K which is colder than the temperatures of the above-mentioned undetected high extinction cores (15.5~K and 15.9~K), none of which are known to contain embedded sources \citep{for09}. There is a tentative trend in HCO$^+$ and HCN to lower extinction-normalized emission with lower temperatures. If the lines are optically thick, as indicated in some cases (see above), then the emission would mainly trace the temperature, explaining this correlation. The detections in HCO$^+$, HCN, and HNC span the full temperature range. While H$^{13}$CO$^+$ and H$^{13}$CN are preferentially detected at lower temperatures, that is mostly a column density effect since colder cores have higher column densities (Fig.~\ref{fig_TvsAV}).

Our finding that only the coldest cores are detected in N$_2$H$^+$ corroborates  a recent similar indication of a correlation between the modeled central temperature of Taurus cores and the presence of N$_2$H$^+$ emission: \cite{mar14} found that the seven N$_2$H$^+$ detections in their sample (from \citealp{hac13}) include four out of the five coldest cores, and they note that N$_2$H$^+$ detections thus seem to be preferentially associated with the coldest cores. Here, we find a direct correlation with the more easily observable effective dust temperature that is not as model-dependent as the central core temperature; it is simply based on a modified blackbody fit to the submillimeter spectral energy distribution.

\section{Summary and Conclusions}

We have observed 52 low-mass molecular cores in the Pipe Nebula in the ground rotational transitions of HCN, H$^{13}$CN, HCO$^{+}$, H$^{13}$CO$^{+}$, HNC, and N$_2$H$^{+}$. These molecular-line observations are interpreted in the context of the dust column densities and effective dust temperatures of the cores, as derived from extinction maps and {\it Herschel} submillimeter mapping, respectively. The column densities and effective dust temperatures toward the cores are generally anti-correlated, i.e., the cores with the highest levels of extinction also have the lowest dust temperatures. This is consistent with an interpretation that the highest-extinction cores indeed contain both cold and dense material. Also, a comparison of effective dust temperature and NH$_3$ gas kinetic temperatures for a subset of our sample demonstrates that the two quantities are correlated, indicating that the effective dust temperature is dominated by dust that is thermodynamically coupled to dense gas, even if the effective dust temperature is elevated by the inclusion of hotter dust in the line-of-sight averaging. At least relative statements based on effective dust temperatures are thus physically meaningful.

While all integrated line intensities increase with line of sight extinction, the rate of increase is different for different molecules. HCO$^+$ and HCN emission, normalized with extinction, decreases with increasing extinction, most likely because the lines become optically thick. This is demonstrated by H$^{13}$CO$^+$ and H$^{13}$CN where the emission, normalized with extinction, is roughly constant within the sample, indicative of a constant abundance and optically thin lines. N$_2$H$^+$ emission normalized with extinction increases with extinction indicative of overproduction of this molecule relative to the others toward the densest cores. Based on previous observations and chemical modeling, N$_2$H$^+$ will only become observable in dense regions when the temperature is low enough for CO to freeze out (T$\sim$20~K; \citealp{bis06}). Supporting this theory, N$_2$H$^+$ emission (absolute and normalized) is generally anti-correlated with the effective dust temperature, to a degree that it is only detected toward the coldest cores in the sample. However, we found two starless cores in our sample with comparable column densities but no N$_2$H$^+$ detections (35 and 87), and these cores are significantly warmer than the ones that do have N$_2$H$^+$ detections. Here, the column density may be due at least partially to chance projections and superpositions of otherwise unrelated material, even though in this case the column densities are higher than the minimum threshold for the detection of N$_2$H$^+$ in cores found by \citet{ber01} and reported as $A_V=4$~mag.

In summary, of common core tracers, N$_2$H$^+$ is the only species to exclusively trace the coldest and densest cores, while the widely detected transitions of HCN, HCO$^+$, HNC and their isotopologues can be used to obtain a census of the total core population. The routine availability of {\it Herschel} effective dust temperatures for large areas of nearby star-forming regions is now providing an excellent starting point for targeted molecular-line studies of the coldest and densest cores.

\begin{acknowledgements}
We would like to thank Ali Ahmad Khostovan who worked with us on this project as a summer student in the SAO REU Summer Intern Program in 2011, an anonymous referee for comments that helped to improve the paper, and the editor, Malcolm  Walmsley, for a careful reading of the manuscript and additional constructive suggestions for improvement. The Mopra radio telescope is part of the Australia Telescope National Facility which is funded by the Commonwealth of Australia for operation as a National Facility managed by CSIRO. The University of New South Wales Digital Filter Bank used for the observations with the Mopra Telescope was provided with support from the Australian Research Council. Herschel is an ESA space observatory with science instruments provided by European-led Principal Investigator consortia and with important participation from NASA. This publication is supported by the Austrian Science Fund (FWF).
\end{acknowledgements}

\bibliography{pipe} 

%

\end{document}